\documentclass[a4paper,USenglish,cleveref,autoref]{lipics-v2021}

\usepackage{amsmath,amssymb}
\usepackage{tikz}
\usetikzlibrary{intersections,positioning,decorations.pathreplacing}
\usepackage{booktabs}
\usepackage{algorithm}
\usepackage[noend]{algpseudocode}
\crefname{observation}{Observation}{Observations}
\Crefname{observation}{Observation}{Observations}
\crefname{appendix}{Appendix}{Appendices}
\Crefname{appendix}{Appendix}{Appendices}

\newcommand{\R}{\mathbb{R}}
\newcommand{\E}{\mathbb{E}}
\newcommand{\Hset}{\mathcal{H}}
\newcommand{\Fam}{\mathcal{F}}
\newcommand{\conv}{\operatorname{conv}}
\newcommand{\eps}{\varepsilon}
\newcommand{\ff}[2]{(#1)_{#2}}          
\newcommand{\lnp}{\ln^{+}}
\newcommand{\Ev}{\mathcal{E}}
\newcommand{\EvT}{\widetilde{\mathcal{E}}}
\newcommand{\Radon}{\operatorname{Radon}}
\newcommand{\Otilde}{\widetilde{O}}

\title{Centerpoints with Few Radon Points}
\titlerunning{Centerpoints with Few Radon Points}

\author{Guilherme D. da Fonseca}{Aix-Marseille Universit\'e and LIS, France \and \url{https://pageperso.lis-lab.fr/guilherme.fonseca/}}{guilherme.fonseca@lis-lab.fr}{https://orcid.org/0000-0002-9807-028X}{}
\author{Kunal Dutta}{Institute of Informatics, University of Warsaw, Poland \and \url{https://www.mimuw.edu.pl/~kdutta/index.html}}{kdutta@mimuw.edu.pl}{https://orcid.org/0000-0003-3055-9326}{Supported by the Polish NCN Opus grant nr. 2023/51/B/ST6/02989.}
\authorrunning{G.\,D. da Fonseca and K. Dutta}
\Copyright{Guilherme D. da Fonseca and Kunal Dutta}

\ccsdesc[500]{Theory of computation~Computational geometry}
\ccsdesc[300]{Theory of computation~Randomness, geometry and discrete structures}

\keywords{Centerpoints, Tukey depth, Radon points, randomized algorithms, high dimensions}

\category{}
\relatedversion{}
\supplement{}
\acknowledgements{}

\EventEditors{}
\EventNoEds{0}
\EventLongTitle{}
\EventShortTitle{}
\EventAcronym{}
\EventYear{}
\EventDate{}
\EventLocation{}
\EventLogo{}
\SeriesVolume{}
\ArticleNo{}
\hideLIPIcs

\begin{document}
\nolinenumbers

\maketitle

\begin{abstract}
  A point $z\in\R^d$ is a $\beta$-centerpoint of an $n$-point set $P\subset\R^d$ if every closed halfspace containing $z$ contains at least $\beta n$ points of~$P$.
  Iterated Radon points yield an $\Omega(1/d^2)$-centerpoint in $\Otilde(d^9)$ time: each round replaces a population by Radon points of random $(d+2)$-tuples.
  Har-Peled and Jones improved the running time to $\Otilde(d^7)$.
  The known analyses certify that \emph{every} point of the population is central.

  We show that $O(\log^2 d)$ rounds of Radon computations suffice if a \emph{single} Radon point of the last round is output.
  The analysis tracks, for every $k$, the probability that $k$ random points of a round lie in a common ``bad'' halfspace, so the union bound over halfspaces is taken only once, for the input sample.
  For points in general position and constant $\eps>0$, a $\frac{1-\eps}{4(d+1)(d+2)}$-centerpoint is computed with constant probability using $O(d^3\log^3 d)$ Radon computations, in $O(d^{3+\omega}\log^3 d)\subseteq O(d^{5.372})$ time ($\omega$: matrix multiplication exponent), or $O(d^6\log^3 d)$ time with Gaussian elimination.
  The algorithm uses only linear algebra; among strongly polynomial algorithms for centerpoints of quality $1/\mathrm{poly}(d)$ it is the fastest known.
  For integer input, grid rounding keeps all points polynomially long and still yields an exact centerpoint.

  Dutta and Pisula obtained quality $\Omega(1/(d\log d\log\log d))$ in weakly polynomial time by solving $\Otilde(d^3)$ linear programs, using Tverberg points with $\Theta(\log d)$ parts instead of Radon points.
  Our analysis extends to rounds of such Tverberg points: for inputs of more than $\Otilde(d)$ points, it gives the same quality from $O(d^2\log^4 d\log\log d)$ linear programs in expectation.
\end{abstract}

\section{Introduction}\label{sec:intro}

Let $P$ be a set of $n$ points in $\R^d$.
A point $z\in\R^d$ is a \emph{$\beta$-centerpoint} of $P$ if every closed halfspace containing $z$ contains at least $\beta n$ points of~$P$; equivalently, $z$ has Tukey depth at least $\beta n$ \cite{t-mpds-75}.
The centerpoint theorem~\cite{r-tgm-46} guarantees the existence of a $\frac{1}{d+1}$-centerpoint, and centerpoints are a basic tool for divide-and-conquer algorithms, geometric separators~\cite{mttv-spnng-97}, and robust statistics~\cite{dg-bpmle-92}.
Computing a $\frac{1}{d+1}$-centerpoint takes $n^{O(d)}$ time with known methods~\cite{cemst-acpir-96}; linear time is known in the plane~\cite{jm-ccfps-94}, and Chan gave an optimal randomized algorithm for the maximum Tukey depth in low dimensions~\cite{c-oramt-04}.
For a long time, the best quality achievable in time polynomial in~$d$ was $\Theta(1/d^2)$, by iterated Radon points (see below).
Cherapanamjeri~\cite{c-cacpt-24} computed, in time polynomial in~$d$ (of high degree), a point close to a point of depth $\Omega(1/d)$, which is a weaker notion of approximation.
Dutta and Pisula~\cite{dp-ncptad-26,dp-centerpoints-arxiv-26}\footnote{The SODA version~\cite{dp-ncptad-26} states a slightly 
weaker bound of $\Omega\left(\frac{1}{d\log^2 d}\right)$; the more accurate bound of $\Omega\left(\frac{1}{d\log d\log\log d}\right)$ is stated in the full version~\cite{dp-centerpoints-arxiv-26}.} computed an $\Omega(1/(d\log d\log\log d))$-centerpoint in weakly polynomial time, by solving $\Otilde(d^3)$ linear programs: they replace Radon points by Tverberg points with $\Theta(\log d)$ parts of random subsets of size $\Otilde(d)$, which they obtain from a random partition by solving a linear program.
Their running time depends on the bit length of the input (through the linear programs), and the only strongly polynomial version they give is subexponential in~$d$.

\subparagraph*{Iterated Radon points.}
Any $d+2$ points of $\R^d$ can be partitioned into two parts whose convex hulls intersect (Radon's theorem~\cite{r-mkmp-21}); a point in the intersection is a \emph{Radon point}.
A Radon point is ``central'' among its $d+2$ defining points: a halfspace containing it must contain at least two of them.
The algorithm of~\cite{cemst-acpir-96} exploits this fact by iterating: replace random groups of $d+2$ points by their Radon points, and repeat.
For $d=1$ this is the iterated median of three, which is closely related to Valiant's construction of monotone formulas for majority~\cite{v-smfm-84} and to the layered random circuits of Hoory, Magen, and Pitassi~\cite{hmp-mcmf-06}.
A full $(d+2)$-ary tree of Radon computations of height $h$ fails to be central with respect to a fixed halfspace with probability doubly exponentially small in~$h$, but a union bound over the $n^{d}$ relevant halfspaces requires $h=\Theta(\log(d\log n))$ and thus a number of leaves that is quasi-polynomial in $d$.
To obtain a polynomial algorithm, the authors of~\cite{cemst-acpir-96} replaced the tree by a layered structure in which every round replaces a whole population of $n$ points by $n$ Radon points of random groups, and showed that for $n=\Omega(d^{4+\delta})$, after $\Theta(d+\log\log n)$ rounds \emph{every} point of the population is an $\Omega(1/d^2)$-centerpoint with high probability; combined with random sampling, this gives an $\Otilde(d^9)$-time algorithm, with $\Otilde(d^6)$ Radon computations of $O(d^3)$ time each, computing a $\frac{2}{3e(d+1)(d+2)}$-centerpoint.
Har-Peled and Jones~\cite{hj-jcps-19,hj-jcps-21} revisited the process: they take a sample of $N=\Theta(d^3\log d)$ points, which is a relative approximation for halfspaces, and replace rounds by an iterative process, analyzed as a random walk (``Radon's urn''), in which every step inserts one Radon point and deletes a random point.
For an accuracy parameter $\eps\in(0,1)$ and a failure probability $\varphi\in(0,1)$, after $O(dN\log N\log(N/\varphi))$ steps of $O(d^3)$ time each, which is $O(d^4\log^3 d)$ steps for constant $\eps$ and $\varphi$, every point of the population is a $\frac{1-\eps}{(d+2)^2}$-centerpoint with probability at least $1-\varphi$; this gives an $\Otilde(d^7)$-time algorithm.
Miller and Sheehy~\cite{ms-acp-10} derandomized the approach, and Har-Peled and Zhou~\cite{hz-iaatp-21} used it for approximate Tverberg partitions.

The analyses of~\cite{cemst-acpir-96,hj-jcps-21} certify that \emph{every} point of the population is central: for each of the $e^{\Lambda}$ relevant ``bad'' halfspaces~$h$, where $\Lambda=\Theta(d\log N)$, the number of population points in $h$ must reach zero with probability $1-e^{-\Lambda}$.
In the round-based process, a pair of points in $h$ reproduces itself with probability polynomially small in~$d$, so the last bad points disappear only after about $\Lambda/\log d=\Theta(d)$ rounds.
In the iterative process, the last bad point disappears only when deleted, at rate $1/N$ per step, hence $O(\Lambda N\log N)$ steps.

\subparagraph*{Our contribution.}
We use a layered structure as in~\cite{cemst-acpir-96} but with only $O(\log^2 d)$ layers (for constant $\eps$ and $\varphi$), and we output a \emph{single} Radon point of the last layer.
We do not attempt to certify that the whole population is central.
Instead, for each layer~$i$ and each $k$ we bound
\[
  A_i(k) \;\approx\; \Pr\bigl[\,k \text{ random points of layer } i \text{ lie in a common bad halfspace}\,\bigr].
\]
A $k$-tuple of layer $i+1$ lies in a bad halfspace $h$ only if each of the $k$ Radon points has two of its inputs in $h$; the $2k$ inputs form a random subset of layer $i$ of size $2k$ minus the number of \emph{collisions}.
This gives a recursion from $A_i$ to $A_{i+1}$ (\cref{lem:recursion}).
For this quantity, the union bound over halfspaces is used only for the input sample, where $k$ distinct sample points lie in a common bad halfspace with probability at most $e^{\Lambda}\beta^k$.
(We also use a union bound over halfspaces to keep the number of points of every layer in every bad halfspace below $R=\Theta(\Lambda)$ (below $2R$ in the first layer); this follows from a simple Chernoff concentration bound that holds from the first layer on.)
Collisions are the main difficulty: they make it cheaper for the adversary to keep a tuple inside a halfspace, and they are what makes certifying the whole population expensive.
We show that they are harmless for the single output point by a simple bound on the moment generating function of the number of distinct indices among $2k$ random ones (\cref{lem:collisions}).
The resulting recursion has the following behavior: the ``threshold'' $\kappa_i$, below which the bound $A_i(k)\le e^{\mu_i(\kappa_i-k)}$ is trivial, decreases geometrically, by a factor $1-\Theta(1/\log d)$ per layer, from $\Theta(d)$ to below~$1$ (\cref{lem:main-phase}), after which a constant number of layers suffices (\cref{lem:final-phase}).

\begin{theorem}\label{thm:intro}
  Let $P$ be a set of $n$ points in general position in $\R^d$, $d\ge2$, and let $\eps,\varphi\in(0,1)$ be constants.
  One can compute a point that is a $\frac{1-\eps}{4(d+1)(d+2)}$-centerpoint of $P$ with probability at least $1-\varphi$ by computing $O(d^3\log^3 d)$ Radon points, in $O(d^{3+\omega}\log^3 d)$ time, where $\omega$ is the matrix multiplication exponent.
  Using Gaussian elimination instead of fast matrix multiplication, the running time is $O(d^6\log^3 d)$.
\end{theorem}

For comparison, the algorithm of Har-Peled and Jones computes $O(d^4\log^3 d)$ Radon points and runs in $O(d^7\log^3 d)$ time, and it guarantees the better quality $\frac{1-\eps}{(d+2)^2}$; the quality guaranteed by \cref{thm:intro} is about four times worse, and slightly better than the $\frac{2}{3e(d+1)(d+2)}$ of~\cite{cemst-acpir-96}.
With the best published bound $\omega<2.371339$~\cite{adwxxz-maffmm-25}, the running time of \cref{thm:intro} is $O(d^{5.372})$.
The dependence on $\eps$ and $\varphi$ is given in \cref{thm:main}: the number of layers grows by an additive $O(\log(1/\varphi)+\log d\,\lnp\log\frac1\eps)$, and apart from drawing the sample, the running time depends only polylogarithmically on $1/\eps$.
\Cref{tab:comparison} compares \cref{thm:intro} with previous algorithms.
The quality guaranteed by \cref{thm:intro} is worse than that of Dutta and Pisula~\cite{dp-ncptad-26,dp-centerpoints-arxiv-26} by a factor of $\Theta(d/(\log d\log\log d))$; on the other hand, the algorithm of \cref{thm:intro} does not solve linear programs and is strongly polynomial.\footnote{\label{fn:strong}We call an algorithm \emph{strongly polynomial} if, in the real RAM, the number of arithmetic operations is polynomial in~$d$ (after sampling) and independent of the bit length of the input.
The Turing-machine notion additionally requires numbers of polynomial bit length; exact iterated Radon computations do not have this property, for our algorithm or for~\cite{cemst-acpir-96,hj-jcps-21}.}
\cref{thm:intro-tverberg} below matches the quality of~\cite{dp-ncptad-26}, but solves linear programs and is weakly polynomial.

\subparagraph*{Layers of Tverberg points.}
The analysis uses only one property of a Radon point: a halfspace containing it contains two of its inputs.
Dutta and Pisula~\cite{dp-ncptad-26,dp-centerpoints-arxiv-26} compute points with the stronger property that a halfspace containing them contains $s$ of their $t$ inputs.
To do so, they solve a linear program for a random $s$-coloring of the inputs, using a lifting argument of Sarkaria~\cite{s-ttnf-92} and Sober\'on~\cite{s-tpwen-19}.
In \cref{sec:tverberg} we run our layered analysis with these \emph{Tverberg gates}.
The constant $\binom t2$ of the recursion becomes $\binom ts$, and the pairs of inputs become $s$-subsets.
Two things improve.
With $s=\Theta(\log d)$ and $t=\Theta(s^2d\log s)$ the quality is $\Theta(s/t)$.
Moreover, the threshold $\kappa_i$ now halves in every layer, so $O(\log d)$ layers suffice.
Together, this gives quality $\Theta(1/(d\log d\log\log d))$ from $O(d^2\log^4d\log\log d)$ gates.

\begin{theorem}\label{thm:intro-tverberg}
  Let $P$ be a set of $n$ points in general position in $\R^d$, let $\eps,\varphi\in(0,1)$ be constants, and assume that $n\ge t$, where $t=O(d\log^2d\log\log d)$ is the number of inputs of a gate.
  One can compute a point that is an $\Omega\bigl(\frac{1}{d\log d\log\log d}\bigr)$-centerpoint of $P$ with probability at least $1-\varphi$ by solving $O(d^2\log^4 d\log\log d)$ linear programs in expectation.
  Each is a feasibility problem with $O(d\log^2 d\log\log d)$ nonnegative variables and $O(d\log d)$ equality constraints.
\end{theorem}

The algorithm of Dutta and Pisula solves $\Theta(d^3\,\mathrm{polylog}\,d)$ linear programs of the same kind and guarantees the same quality.
With a solver taking $O_w(d^{2.5}\,\mathrm{polylog}\,d)$ time per program, they state the bound $O_w(d^{5.5}\,\mathrm{polylog}\,d)$, where $O_w$ indicates a weakly polynomial bound.
\Cref{thm:intro-tverberg} combines their subroutine with our analysis.

\begin{table}[t]
  \centering
  \caption{Centerpoint algorithms polynomial in $d$ (constant $\eps$, $\varphi$).
    Best known times use $T_{\mathrm{Radon}}=O(d^\omega)$ (\cref{lem:radon}).
    $T_{\mathrm{LP}}$ is the time to solve a linear program with $\Otilde(d)$ variables and constraints exactly; it depends on the bit length, which differs between the two LP-based algorithms, so we do not compare their times.
    \emph{Strong}: strongly polynomial in the sense of footnote~\ref{fn:strong}.}
  \label{tab:comparison}
  \begin{tabular}{@{}lllll@{}}
    \toprule
    Algorithm & Quality & Time & Best known time & Strong \\
    \midrule
    \cite{cemst-acpir-96} & $\frac{2}{3e(d+1)(d+2)}$ & $\Otilde(d^6)\cdot T_{\mathrm{Radon}}$ & $\Otilde(d^{8.372})$ & yes \\[2pt]
    Har-Peled, Jones~\cite{hj-jcps-21} & $\frac{1-\eps}{(d+2)^2}$ & $\Otilde(d^4)\cdot T_{\mathrm{Radon}}$ & $\Otilde(d^{6.372})$ & yes \\[2pt]
    Cherapanamjeri~\cite{c-cacpt-24} & $\Omega(\frac1d)$, weaker notion & $\mathrm{poly}(d)$ & $\mathrm{poly}(d)$ & no \\[2pt]
    Dutta, Pisula~\cite{dp-ncptad-26,dp-centerpoints-arxiv-26} & $\Omega\bigl(\frac{1}{d\log d\log\log d}\bigr)$ & $\Otilde(d^3)\cdot T_{\mathrm{LP}}$ & -- & no \\[2pt]
    \Cref{thm:intro} & $\frac{1-\eps}{4(d+1)(d+2)}$ & $\Otilde(d^3)\cdot T_{\mathrm{Radon}}$ & $\Otilde(d^{5.372})$ & yes \\[2pt]
    \Cref{thm:intro-tverberg} & $\Omega\bigl(\frac{1}{d\log d\log\log d}\bigr)$ & $\Otilde(d^2)\cdot T_{\mathrm{LP}}$ & -- & no \\
    \bottomrule
  \end{tabular}
\end{table}

\subparagraph*{Remarks.}
(i) Replacing Gaussian elimination by fast matrix multiplication in the algorithm of Har-Peled and Jones alone gives $\Otilde(d^{4+\omega})$; our improvement in the number of Radon computations is independent of this.
(ii) In exact arithmetic, the bit length of the points grows geometrically with the number of rounds~\cite[Section~9]{cemst-acpir-96}.
For integer input coordinates of $B$ bits, $n$ at least the sample size, and $\eps\le\frac13$, rounding every Radon (or Tverberg) point to a grid of mesh $2^{-O(d(B+\log d))}$ affects the analysis only through a loss of $d$ points per bad halfspace.
Hence the output is still an exact centerpoint, of quality $(1-\eps)^2\beta$ instead of $(1-\eps)\beta$, and all computed points have $O(d(B+\log d))$ bits per coordinate.
(iii) A natural idea to speed up iterated Radon computations is to reuse a factored matrix of $d+1$ points to compute Radon points with many different last points in $O(d^2)$ time each.
However, Radon points computed from a common base are strongly correlated, and we do not see how to exploit such sharing.

\subparagraph*{Organization.}
\Cref{sec:prelim} collects preliminaries.
\Cref{sec:algorithm} describes and analyzes the algorithm based on Radon points, which proves \cref{thm:intro}.
\Cref{sec:tverberg} extends the analysis to layers of Tverberg points and proves \cref{thm:intro-tverberg}.
\Cref{sec:bits} contains the analysis and discussion on bit-length complexities.

\section{Preliminaries}\label{sec:prelim}

Throughout, $d\ge 2$, $t:=d+2$, and $c:=\binom{t}{2}=\frac{(d+1)(d+2)}{2}\ge 6$, except in \cref{sec:tverberg}, where $t$ is redefined.
We write $\ff{x}{k}=x(x-1)\cdots(x-k+1)$ for the falling factorial and $\lnp x=\max\{0,\ln x\}$.
A \emph{halfspace} is a set $\{x: \langle a,x\rangle > b\}$ (open) or $\{x:\langle a,x\rangle\ge b\}$ (closed) with $a\ne 0$.
We work in the real RAM model, and we assume that $P$ is given as an array, so that a uniformly random point of $P$ can be drawn in constant time.
Point sequences $Q=(q_1,\dots,q_m)$ may contain repeated points; subsets of a sequence are always subsets of its \emph{index} set $[m]=\{1,\dots,m\}$, and for $J\subseteq[m]$ we write $Q[J]=\{q_j:j\in J\}$.

\subsection{Radon Points}

\begin{lemma}\label{lem:radon}
  Let $X=(x_1,\dots,x_t)$ be a sequence of $t=d+2$ points of $\R^d$.
  In $O(d^\omega)$ time (or $O(d^3)$ time by Gaussian elimination) one can compute a partition $[t]=I^+\cup I^-$ into nonempty sets and a point $z\in \conv X[I^+]\cap\conv X[I^-]$, called a \emph{Radon point} of~$X$.
  Moreover, every halfspace (open or closed) that contains $z$ contains $x_i$ and $x_j$ for some $i\in I^+$ and $j\in I^-$; in particular, it contains the points of $X$ at two distinct indices.
\end{lemma}

\begin{proof}
  Let $M$ be the $(d+1)\times(d+2)$ matrix whose $i$-th column is $(x_i,1)$.
  Since $M$ has more columns than rows, it has a nonzero kernel vector $\lambda$, which can be computed in $O(d^\omega)$ time from an LSP decomposition~\cite{imh-gflmd-82} (see also~\cite{bh-tfifmm-74}), or in $O(d^3)$ time by Gaussian elimination.
  Since $\sum_i\lambda_i=0$ and $\lambda\ne0$, the set $I^+=\{i:\lambda_i>0\}$ is nonempty and so is $\{i:\lambda_i<0\}$; let $I^-=[t]\setminus I^+$ and $\sigma=\sum_{i\in I^+}\lambda_i>0$.
  From $\sum_i\lambda_ix_i=0$ we get $z:=\sum_{i\in I^+}\frac{\lambda_i}{\sigma}x_i=\sum_{i\in I^-}\frac{-\lambda_i}{\sigma}x_i$, a convex combination of $X[I^+]$ and of $X[I^-]$ (the coefficients $-\lambda_i/\sigma$, $i\in I^-$, are nonnegative and sum to~$1$).
  If a halfspace $h$ contains no $x_i$ with $i\in I^+$, then the complement of $h$, which is convex, contains $X[I^+]$ and hence~$z$.
  The same argument applies to $I^-$ (the indices with $\lambda_i=0$ do not contribute to the combination).
\end{proof}

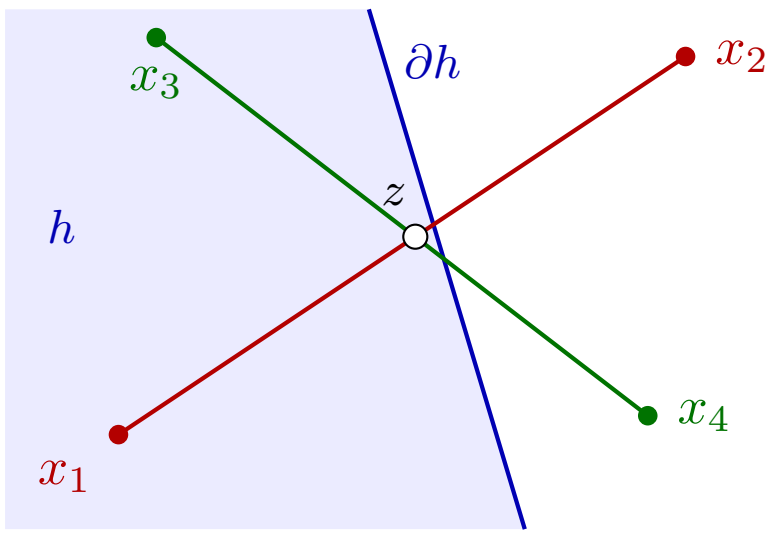
\begin{figure}[t]
  \centering
  \begin{tikzpicture}[scale=1.25, pt/.style={circle,fill,inner sep=1.3pt}]
    \coordinate (p1) at (0,0);
    \coordinate (p2) at (3,2);
    \coordinate (p3) at (0.2,2.1);
    \coordinate (p4) at (2.8,0.1);
    \fill[blue!8] (-0.6,-0.5) -- (2.15,-0.5) -- (1.325,2.25) -- (-0.6,2.25) -- cycle;
    \draw[blue!70!black, thick] (2.15,-0.5) -- (1.325,2.25) node[pos=0.9, right, font=\small] {$\partial h$};
    \node[blue!70!black, font=\small] at (-0.3,1.1) {$h$};
    \draw[red!70!black, thick, name path=A] (p1) -- (p2);
    \draw[green!45!black, thick, name path=B] (p3) -- (p4);
    \path[name intersections={of=A and B, by=z}];
    \node[pt, red!70!black, label={[red!70!black]below left:$x_1$}] at (p1) {};
    \node[pt, red!70!black, label={[red!70!black]right:$x_2$}] at (p2) {};
    \node[pt, green!45!black, label={[green!45!black]below:$x_3$}] at (p3) {};
    \node[pt, green!45!black, label={[green!45!black]right:$x_4$}] at (p4) {};
    \node[draw, fill=white, circle, inner sep=1.6pt, label={[font=\small,xshift=-4pt]above:$z$}] at (z) {};
  \end{tikzpicture}
  \caption{A Radon point $z$; a halfspace $h\ni z$ contains a point of each part.}
  \label{fig:radon}
\end{figure}

We write $\Radon(X)$ for the Radon point computed by \cref{lem:radon}; see \cref{fig:radon}.

\subsection{Centerpoints and Bad Halfspaces}

For a finite set $S\subset\R^d$ and an integer $k\ge1$, a point $z$ is a \emph{$k/|S|$-centerpoint} of $S$ if every closed halfspace containing $z$ contains at least $k$ points of $S$.
The following lemma reduces centrality with respect to $S$ to avoiding a finite family of open halfspaces.
It is essentially Theorem~2.3 of~\cite{cemst-acpir-96}, due to Danzer, Gr\"unbaum, and Klee~\cite{dgk-htr-63}; we prove it for completeness.
A set $S\subset\R^d$ is in \emph{general position} if no $d+1$ of its points lie on a hyperplane.

\begin{lemma}\label{lem:bad-family}
  Let $S\subset\R^d$ be a set of $\nu$ points in general position and let $1\le k_0\le \nu-d$.
  Let $\Hset$ be the family of open halfspaces $h$ whose boundary hyperplane passes through $d$ points of $S$ and such that $|h\cap S|\le k_0-1$.
  Then $|\Hset|\le 2\binom{\nu}{d}$, and every point $z\notin\bigcup\Hset$ is a $k_0/\nu$-centerpoint of~$S$.
\end{lemma}

\begin{proof}
  Every $d$ points of $S$ span a unique hyperplane, which bounds two open halfspaces, so $|\Hset|\le 2\binom{\nu}{d}$.
  Suppose $z$ is not a $k_0/\nu$-centerpoint of $S$: there is a closed halfspace $h\ni z$ with $T:=h\cap S$ of size at most $k_0-1$.
  The complement of $h$ is convex and contains $S\setminus T$, so $z\notin\conv(S\setminus T)$.
  Since $|S\setminus T|\ge \nu-k_0+1\ge d+1$ and $S$ is in general position, $\conv(S\setminus T)$ is a full-dimensional polytope, and it is the intersection of its facet-defining closed halfspaces.
  Hence some facet $F$ has $z$ strictly on its outer side.
  By general position the hyperplane of $F$ contains exactly $d$ points of $S$, and the open outer halfspace $h^*$ of $F$ contains no point of $S\setminus T$.
  Hence $|h^*\cap S|\le|T|\le k_0-1$, so $h^*\in\Hset$ and $z\in h^*$.
\end{proof}

We call the members of $\Hset$ \emph{bad halfspaces}.

\subsection{Relative Approximations and Tail Bounds}

Let $P$ be a set of $n$ points.
A subset $S\subseteq P$ is a \emph{$(\rho,\eps)$-relative approximation} for halfspaces if for every halfspace $h$, $\bigl|\frac{|h\cap S|}{|S|}-\frac{|h\cap P|}{n}\bigr|\le\eps\max\bigl\{\frac{|h\cap P|}{n},\rho\bigr\}$; the same definition applies to a sequence $S$ of points of $P$, counted with multiplicity.
Since the range space of halfspaces has VC-dimension $d+1$~\cite{m-ldg-02}, a sequence of
\begin{equation}\label{eq:n0}
  n_0 = \Bigl\lceil\frac{C_0}{\eps^{2}\rho}\bigl(d\ln\tfrac1\rho+\ln\tfrac1{\varphi'}\bigr)\Bigr\rceil
\end{equation}
independent uniformly random points of $P$ is a $(\rho,\eps)$-relative approximation with probability at least $1-\varphi'$, for an absolute constant $C_0\ge1$~\cite{lls-ibscl-01,hs-rag-11,h-gaa-11}.
We need a sample without repetitions, which we obtain by rejection.

\begin{lemma}\label{lem:sample}
  Let $n\ge n_0^2$ and let $Y$ be a sequence of $n_0$ independent uniformly random points of $P$.
  Then $Y$ has no repeated point with probability at least $\frac12$, and conditioned on this event, $Y$ is a uniformly random $n_0$-subset of $P$ and a $(\rho,\eps)$-relative approximation with probability at least $1-2\varphi'$.
\end{lemma}

\begin{proof}
  The probability of a repetition is at most $\binom{n_0}{2}/n\le\frac12$.
  Conditioned on no repetition, all $n_0$-subsets are equally likely by symmetry, and the probability of not being a relative approximation is at most $\varphi'/\frac12$.
\end{proof}

\begin{lemma}\label{lem:transfer}
  Let $S$ be a $(\rho,\eps)$-relative approximation of $P$ for halfspaces and let $z$ be a $\beta$-centerpoint of $S$ with $\beta\ge(1+\eps)\rho$.
  Then $z$ is a $(1-\eps)\beta$-centerpoint of~$P$.
\end{lemma}

\begin{proof}
  Let $h\ni z$ be a closed halfspace and $x=|h\cap P|/n$.
  Then $\beta\le|h\cap S|/|S|\le x+\eps\max\{x,\rho\}$.
  If $x<\rho$, this gives $\beta<(1+\eps)\rho$, a contradiction.
  Hence $x\ge\rho$ and $\beta\le(1+\eps)x$, so $x\ge\beta/(1+\eps)\ge(1-\eps)\beta$.
\end{proof}

\begin{lemma}[Chernoff bound, see~\cite{mu-pc-05}]\label{lem:chernoff}
  Let $X$ be a sum of independent indicator variables with $\E[X]\le\mu$, and let $a\ge\mu$, $a>0$.
  Then $\Pr[X\ge a]\le(e\mu/a)^a$.
\end{lemma}

\begin{proof}
  If $\mu=0$ then $X=0$ and the claim is trivial.
  Otherwise, for $\vartheta=\ln(a/\mu)\ge0$ we have $\E[e^{\vartheta X}]\le e^{\mu(e^\vartheta-1)}$, so by Markov's inequality $\Pr[X\ge a]\le e^{-\vartheta a+\mu(e^\vartheta-1)}=(\mu/a)^ae^{a-\mu}\le(e\mu/a)^a$.
\end{proof}

\section{Layers of Radon Points}\label{sec:algorithm}

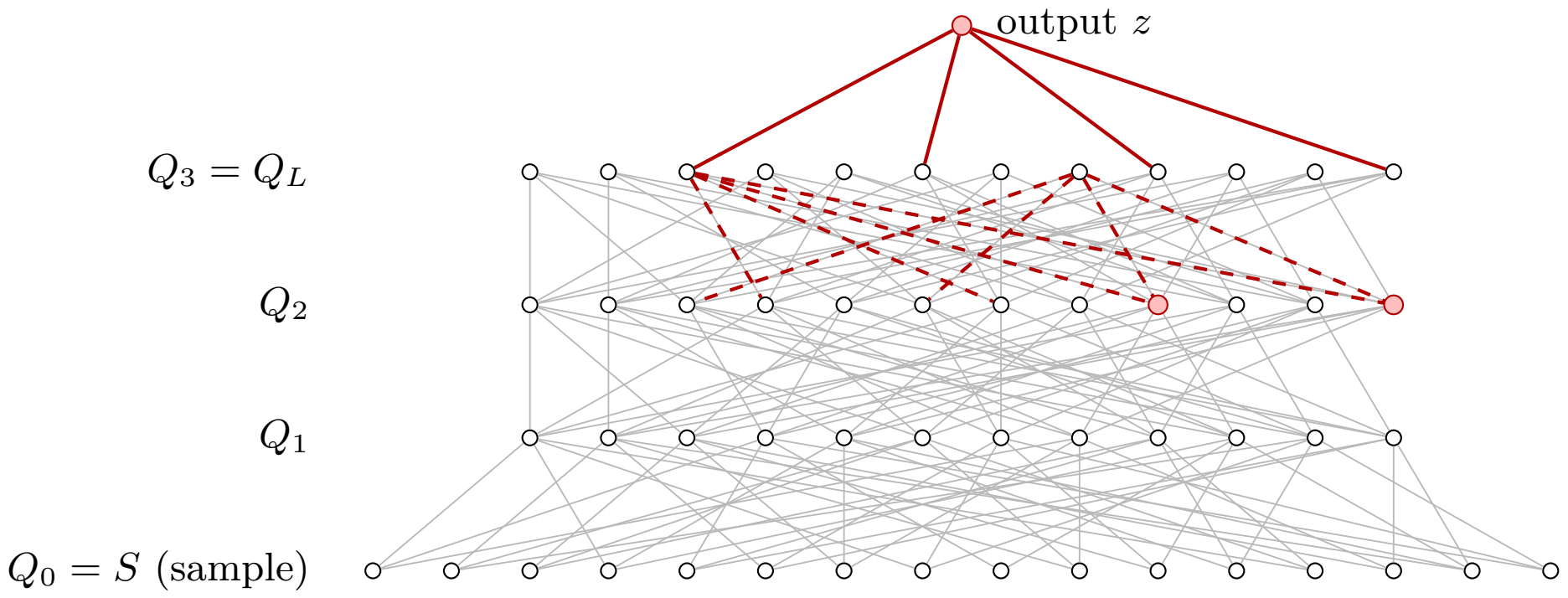
\begin{figure}[t]
  \centering
  \begin{tikzpicture}[x=0.62cm,y=1.05cm,
      nd/.style={circle,draw,fill=white,inner sep=1.2pt},
      hl/.style={circle,draw=red!70!black,fill=red!25,inner sep=1.5pt},
      e/.style={gray!55,line width=0.35pt},
      he/.style={red!70!black,line width=0.8pt}]
    \foreach \x in {0,...,15} { \node[nd] (a\x) at (\x,0) {}; }
    \foreach \l in {1,2,3} { \foreach \x in {0,...,11} { \node[nd] (q\l-\x) at (\x+2,\l) {}; } }
    \node[hl] (out) at (7.5,4.1) {};
    \node[right=2pt of out, font=\small] {output $z$};
    \foreach \c/\ps in {0/{0,3,7,12},1/{1,5,9,14},2/{2,6,10,13},3/{0,4,8,15},4/{3,6,11,14},5/{1,2,9,12},
                         6/{5,7,10,15},7/{0,6,9,13},8/{2,4,11,12},9/{3,8,10,14},10/{1,7,11,15},11/{4,5,13,14}}
      { \foreach \p in \ps { \draw[e] (q1-\c) -- (a\p); } }
    \foreach \l/\lm in {2/1,3/2} {
      \foreach \c/\ps in {0/{0,2,5,9},1/{1,4,7,10},2/{3,6,8,11},3/{0,5,7,11},4/{2,3,9,10},5/{1,6,8,9},
                           6/{0,4,6,10},7/{2,5,8,11},8/{1,3,7,9},9/{0,4,8,10},10/{3,5,6,11},11/{1,2,4,7}}
        { \foreach \p in \ps { \draw[e] (q\l-\c) -- (q\lm-\p); } } }
    \foreach \p in {2,5,8,11} { \draw[he] (out) -- (q3-\p); }
    \foreach \p in {3,6,8,11} { \draw[he,dashed] (q3-2) -- (q2-\p); }
    \foreach \p in {2,5,8,11} { \draw[he,dashed] (q3-7) -- (q2-\p); }
    \node[hl] at (q2-8) {}; \node[hl] at (q2-11) {};
    \foreach \l/\lab in {0/{$Q_0=S$ (sample)},1/{$Q_1$},2/{$Q_2$},3/{$Q_3=Q_L$}}
      { \node[anchor=east, font=\small] at (-0.6,\l) {\lab}; }
  \end{tikzpicture}
  \caption{Layers for $d=2$.
    Two points of $Q_3$ (dashed) share two parents: a collision.}
  \label{fig:layers}
\end{figure}

The algorithm first draws a sample $S$ of $P$, and then builds $L$ \emph{layers} $Q_1,\dots,Q_L$ of $m$ points each: every point of $Q_i$ is the Radon point of $t$ points of $Q_{i-1}$ at distinct random indices, where $Q_0=S$.
The output is the Radon point of $t$ random points of the last layer; see \cref{alg:layered} and \cref{fig:layers}.
The algorithm is the polynomial-time algorithm of~\cite[Algorithm~4]{cemst-acpir-96} with four changes: the number of layers is $O(\log^2 d+\log(1/\varphi)+\log d\,\lnp\log(1/\eps))$ instead of $\Theta(d+\log\log n)$; the layers have size $m=\Theta(d^2 R)$ with $R$ defined below; the $t$ points of a group are chosen at distinct indices (without replacement); and the output is a single Radon point of the last layer instead of an arbitrary point of it.

\begin{algorithm}[t]
  \caption{\textsc{LayeredRadon}$(P,\eps,\varphi)$}\label{alg:layered}
  \begin{algorithmic}[1]
    \State $\beta\gets\frac{1}{4(d+1)(d+2)}$; \ \textbf{if} $n\le1/\beta$ \textbf{then return} any point of $P$
    \State $n_0\gets$ \eqref{eq:n0} with $\rho=\beta/2$ and $\varphi'=\varphi/12$
    \If{$n\le n_0^2$} \ $S\gets P$
    \Else
      \State repeat up to $\lceil\log_2(6/\varphi)\rceil$ times: draw $n_0$ independent uniformly random points of $P$; if they are distinct, let $S$ be these points and stop
      \State \textbf{if} no attempt succeeded \textbf{then return} any point of $P$
    \EndIf
    \State $Q_0\gets S$; compute the parameters $\Lambda,L,R,m$ of \cref{tab:params} from $\nu:=|S|$
    \For{$i=1,\dots,L$}
      \For{$u=1,\dots,m$}
        \State choose a uniformly random sequence $(j_1,\dots,j_t)$ of distinct indices of $Q_{i-1}$
        \State $Q_i[u]\gets\Radon(Q_{i-1}[j_1],\dots,Q_{i-1}[j_t])$
      \EndFor
    \EndFor
    \State choose a uniformly random sequence $(j_1,\dots,j_t)$ of distinct indices of $Q_L$
    \State \Return $z=\Radon(Q_L[j_1],\dots,Q_L[j_t])$
  \end{algorithmic}
\end{algorithm}

\subparagraph*{Parameters.} The parameters are listed in \cref{tab:params}.
The target quality is $\beta:=\frac{1}{8c}=\frac{1}{4(d+1)(d+2)}$.
If $n\le1/\beta$, every point of $P$ is a $\beta$-centerpoint, since every closed halfspace containing it contains at least $1\ge\beta n$ points; hence we assume $n>1/\beta$ from now on.
The sample $S$ has $\nu:=|S|\in\{n_0,n\}$ points, and $\Hset$ is the family of bad halfspaces of \cref{lem:bad-family} for $S$ and $k_0:=\lceil\beta\nu\rceil$.
Each bad halfspace contains at most $k_0-1<\beta\nu$ points of~$S$.
Note that $\Lambda$ is a number computed from $\nu$, not from $\Hset$, and that $|\Hset|\le2\binom{\nu}{d}\le e^{\Lambda}$.
The number of layers $L=1+L_1+L_2$ consists of a first layer, a \emph{main phase} of $L_1$ layers, and a \emph{final phase} of $L_2$ layers.

\begin{table}[t]
  \centering
  \renewcommand{\arraystretch}{1.25}
  \caption{Parameters of \cref{alg:layered}; $t=d+2$ and $c=\binom{t}{2}$.}
  \label{tab:params}
  \begin{tabular}{@{}ll@{}}
    \toprule
    quality & $\beta=\frac{1}{8c}=\frac{1}{4(d+1)(d+2)}$\\
    sample & $\nu=|S|$, \quad $\Lambda:=\ln2+d\ln\frac{e\nu}{d}$\\
    layers & $\kappa_1=\frac{\Lambda}{\ln(64c)}$, \quad $L_1=\bigl\lceil(\ln(4c)+2)\,\lnp\kappa_1\bigr\rceil$,\\
           & $L_2=1+\bigl\lceil\tfrac13\ln\tfrac1\varphi\bigr\rceil$, \quad $L=1+L_1+L_2$\\
    count threshold & $R=\bigl\lceil\tfrac32\bigl(\Lambda+\ln\tfrac{3L}{\varphi}\bigr)\bigr\rceil$\\
    layer size & $m=\lceil 8ecR\rceil$, \quad $x:=\frac{2R}{m}\le\frac{1}{4ec}$, \quad $\ell:=\ln\frac{m}{2R}=\ln\frac1x\ge\ln(4c)+1$\\
    \bottomrule
  \end{tabular}
\end{table}

\subsection{Overview of the Analysis}\label{sec:overview}

\begin{lemma}\label{lem:sample-ok}
  With probability at least $1-\varphi/3$, \cref{alg:layered} either returns in its first line (and then correctly), or $S$ is a $(\beta/2,\eps)$-relative approximation of $P$ consisting of distinct points.
  Moreover, $\nu>1/\beta$ and $k_0\le\nu-d$.
\end{lemma}

\begin{proof}
  If $n\le n_0^2$ then $S=P$ is trivially a relative approximation.
  Otherwise, by \cref{lem:sample}, each attempt succeeds with probability at least $\frac12$, so all $\lceil\log_2(6/\varphi)\rceil$ attempts fail with probability at most $\varphi/6$, and a successful sample is a relative approximation except with probability $2\varphi'=\varphi/6$.
  Since $n>1/\beta$ and $n_0\ge 1/(\eps^2\rho)>1/\beta$, we have $\nu>1/\beta=4(d+1)(d+2)$, hence $k_0\le\beta\nu+1\le\nu-d$, because $(1-\beta)\nu\ge\nu/2\ge d+1$.
\end{proof}

Throughout the rest of this section we condition on the outcome of \cref{lem:sample-ok}, and we assume that $S$ is a $(\beta/2,\eps)$-relative approximation of distinct points.
Since $P$ is in general position, so is $S$.
By \cref{lem:transfer} it suffices to show that the output $z$ is a $\beta$-centerpoint of $S$ with probability at least $1-2\varphi/3$, and by \cref{lem:bad-family} it suffices that $z$ lies in no bad halfspace.
We write $\Fam_i$ for the $\sigma$-algebra generated by $Q_0,\dots,Q_i$; note that $\Hset$ is determined by $\Fam_0$.
General position is used only for $S$; the layers may contain repeated or degenerate points, which is harmless since \cref{lem:radon} holds for arbitrary sequences.

The analysis proceeds as follows.
\Cref{lem:output} reduces the failure probability of the output to $A_L(2)$, the probability that two random points of the last layer lie in a common bad halfspace, and \cref{lem:recursion} bounds $A_{i+1}$ in terms of $A_i$ and the number of collisions, which \cref{lem:collisions} controls.
\Cref{lem:counts} keeps the number of points of every layer in every bad halfspace below the count threshold.
The bound on $A_1$ (\cref{lem:first}) starts the recursion with a bound of the form $e^{\mu(\kappa-k)}$, which is trivial for $k\le\kappa$; in the main phase (\cref{lem:step,lem:main-phase}) the threshold $\kappa$ shrinks geometrically until $\kappa\le1$, at which point the bound is a pure geometric bound in~$k$, and in the final phase (\cref{lem:final-phase}) its base decreases until $A_L(2)$ is small.
\Cref{thm:main} combines these steps.

\subsection{The Tuple Functional}

For a sequence $Q$ of $m'$ points and $1\le k\le m'$, let
\[
  \pi_k(Q) := \Pr_J\bigl[\exists h\in\Hset: Q[J]\subseteq h\bigr],
\]
where $J$ is a uniformly random $k$-subset of $[m']$, and let $\pi_k(Q):=0$ for $k>m'$.
For $h\in\Hset$ let $r_h(Q):=|\{u\in[m']: q_u\in h\}|$ be the number of indices of $Q$ in $h$.

\begin{observation}\label{obs:pi}
  We have $\pi_k(Q)\le1$, $\pi_k(Q)$ is nonincreasing in $k$, and
  \[\pi_k(Q)\le\sum_{h\in\Hset}\frac{\ff{r_h(Q)}{k}}{\ff{m'}{k}}.\]
  In particular, $\pi_k(Q)=0$ if $r_h(Q)<k$ for all $h\in\Hset$.
\end{observation}

\begin{proof}
  A uniformly random $k$-subset is obtained from a uniformly random $(k+1)$-subset by deleting a random element, and if the points of a set lie in a common $h$, so do the points of each of its subsets; this gives monotonicity.
  The last bound is the union bound over $h$, since $\Pr_J[Q[J]\subseteq h]=\binom{r_h(Q)}{k}/\binom{m'}{k}$.
\end{proof}

To control the counts $r_h$, let $\Ev_1$ be the event that $r_h(Q_1)\le 2R$ for all $h\in\Hset$, for $i\ge2$ let $\Ev_i$ be the event that $r_h(Q_i)\le R$ for all $h\in\Hset$, let $\EvT_0$ be the sure event, $\EvT_i:=\Ev_1\cap\dots\cap\Ev_i$, and define
\[
  A_i(k) := \E\bigl[\pi_k(Q_i)\cdot\mathbf 1_{\EvT_i}\bigr]\qquad(1\le i\le L,\ k\ge1).
\]
By \cref{obs:pi}, $A_i(k)\le1$, $A_i$ is nonincreasing, $A_1(k)=0$ for $k>2R$, and $A_i(k)=0$ for $k>R$ if $i\ge2$.

\begin{lemma}[Output]\label{lem:output}
  $\Pr[z\in\bigcup\Hset]\le\Pr[\overline{\EvT_L}]+c\cdot A_L(2)$.
\end{lemma}

\begin{proof}
  Condition on $\Fam_L$.
  If $z\in h$ for some $h\in\Hset$ then, by \cref{lem:radon}, $h$ contains the points of $Q_L$ at two of the $t$ chosen indices.
  Since the chosen sequence of indices is uniformly random, for each of the $c$ pairs of positions the pair of chosen indices is a uniformly random $2$-subset of $[m]$, so the conditional probability is at most $c\,\pi_2(Q_L)$.
  Hence $\Pr[z\in\bigcup\Hset]\le\Pr[\overline{\EvT_L}]+\E[\mathbf 1_{\EvT_L}\,c\,\pi_2(Q_L)]$.
\end{proof}

For $k\ge1$ and a sequence of length $m'\ge2$, let $U_k$ denote the union of $k$ independent uniformly random $2$-subsets of $[m']$, and let $|U_k|$ be its size, the number of distinct indices among the $2k$ chosen ones; $2k-|U_k|$ is the number of \emph{collisions}.

\begin{lemma}[Tuple recursion]\label{lem:recursion}
  For $1\le i<L$ and $1\le k\le m$,
  \[
    \E\bigl[\pi_k(Q_{i+1})\,\big|\,\Fam_i\bigr]\le c^k\sum_{j}\Pr\bigl[|U_k|=j\bigr]\,\pi_j(Q_i),
  \]
  where $U_k$ refers to the index set $[m]$ of $Q_i$.
  Consequently, for all $k\ge1$,
  \[
    A_{i+1}(k)\le\min\Bigl\{1,\;c^k\sum_j\Pr\bigl[|U_k|=j\bigr]\,A_i(j)\Bigr\}.
  \]
\end{lemma}

\begin{figure}[t]
  \centering
  \begin{tikzpicture}[x=0.8cm,y=1.5cm,
      nd/.style={circle,draw,fill=white,inner sep=1.3pt},
      sel/.style={circle,draw=red!70!black,fill=red!25,inner sep=1.6pt},
      e/.style={gray!60,line width=0.4pt},
      he/.style={red!70!black,line width=1pt}]
    \foreach \x in {0,...,13} { \node[nd] (b\x) at (\x,0) {}; }
    \foreach \x/\X in {0/2,1/6.5,2/11} { \node[sel] (u\x) at (\X,1) {}; }
    \node[font=\small, anchor=east, align=right] at (-0.5,1) {$k=3$ points\\of $Q_{i+1}$};
    \node[font=\small, anchor=east] at (-0.5,0) {$Q_i$};
    \foreach \p in {0,3} { \draw[e] (u0) -- (b\p); } \foreach \p in {1,5} { \draw[he] (u0) -- (b\p); }
    \foreach \p in {4,9} { \draw[e] (u1) -- (b\p); } \foreach \p in {5,8} { \draw[he] (u1) -- (b\p); }
    \foreach \p in {7,13} { \draw[e] (u2) -- (b\p); } \foreach \p in {10,12} { \draw[he] (u2) -- (b\p); }
    \foreach \p in {1,5,8,10,12} { \node[sel] at (b\p) {}; }
    \draw[decorate,decoration={brace,mirror,amplitude=5pt}] (0.7,-0.25) -- (12.3,-0.25)
      node[midway,below=6pt,font=\footnotesize,align=center] {$U$: $2k=6$ chosen indices, $|U|=5$ distinct\\(one collision at the shared parent)};
  \end{tikzpicture}
  \caption{\Cref{lem:recursion}: each selected point in $h$ has a pair of parents in $h$ (thick).}
  \label{fig:tuple}
\end{figure}
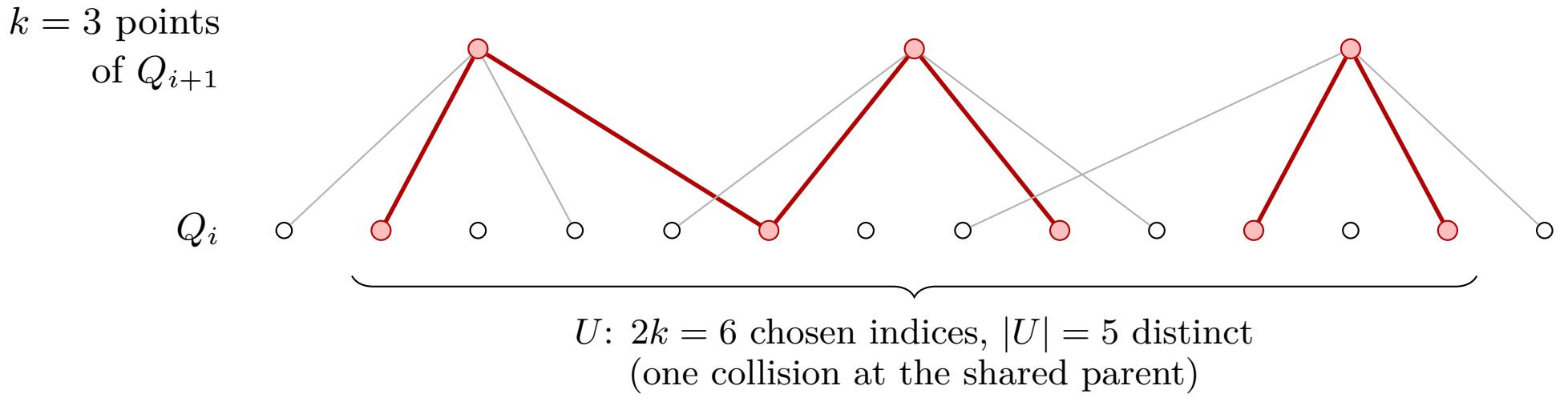

\begin{proof}
  Let $J=\{u_1<\dots<u_k\}$ be a uniformly random $k$-subset of the indices of $Q_{i+1}$.
  For $p\in[k]$, the point $Q_{i+1}[u_p]$ is the Radon point of $Q_i$ at a uniformly random sequence $(\sigma_p(1),\dots,\sigma_p(t))$ of distinct indices, and these sequences are independent of each other, of $J$, and of $\Fam_i$.
  If all points of $Q_{i+1}[J]$ lie in some $h\in\Hset$, then by \cref{lem:radon} for every $p\in[k]$ there are positions $a_p<b_p$ in $[t]$ with $Q_i[\sigma_p(a_p)],Q_i[\sigma_p(b_p)]\in h$.
  We take a union bound over the $c^k$ choices of the positions $(a_p,b_p)_{p\in[k]}$.
  For a fixed choice, the sets $\{\sigma_p(a_p),\sigma_p(b_p)\}$, $p\in[k]$, are independent uniformly random $2$-subsets of the indices of $Q_i$, so their union $U$ is distributed as $U_k$, and since this distribution is invariant under permutations of the indices, conditioned on $|U|=j$, $U$ is a uniformly random $j$-subset, independent of $\Fam_i$.
  The event that all points of $Q_i[U]$ lie in a common bad halfspace therefore has conditional probability $\sum_j\Pr[|U_k|=j]\,\pi_j(Q_i)$, which proves the first claim; see \cref{fig:tuple}.
  For the second claim, if $k>m$ then $A_{i+1}(k)=0$.
  Otherwise, since $\EvT_{i+1}\subseteq\EvT_i$ and $\EvT_i$ is determined by $\Fam_i$, $A_{i+1}(k)\le\E[\mathbf 1_{\EvT_i}\,\E[\pi_k(Q_{i+1})\mid\Fam_i]]$, and the first claim applies.
  Finally, $A_{i+1}(k)\le1$.
\end{proof}

\begin{lemma}[Collisions]\label{lem:collisions}
  Let $m'\ge2$, $k\ge1$, and $w\in(0,1]$.
  Then $\E\bigl[w^{|U_k|}\bigr]\le w^2\bigl(w+\frac{2k}{m'}\bigr)^{2k-2}$.
\end{lemma}

\begin{proof}
  Reveal the $2k$ indices $y_1,\dots,y_{2k}$ one at a time, where $\{y_{2p-1},y_{2p}\}$ is the $p$-th pair; thus $y_{2p-1}$ is uniform in $[m']$ and $y_{2p}$ is uniform in $[m']\setminus\{y_{2p-1}\}$.
  Let $D_q$ be the number of distinct values among $y_1,\dots,y_q$, so $D_2=2$ and $D_q=D_{q-1}+\xi_q$ with $\xi_q\in\{0,1\}$.
  Conditioned on $y_1,\dots,y_{q-1}$, the probability that $\xi_q=0$, i.e., that $y_q$ repeats a value seen so far, is $D_{q-1}/m'$ for odd $q$ and $(D_{q-1}-1)/(m'-1)$ for even $q$; in both cases it is at most $\frac{q-1}{m'}$, since $D_{q-1}\le q-1$ and $\frac{q-2}{m'-1}\le\frac{q-1}{m'}$ for $q\le m'+1$ (for larger $q$ the bound exceeds~$1$).
  Hence $\E[w^{\xi_q}\mid y_1,\dots,y_{q-1}]=w+(1-w)\Pr[\xi_q=0\mid\cdot]\le w+\frac{2k}{m'}$, and multiplying these bounds for $q=3,\dots,2k$ gives the claim.
\end{proof}

\subsection{Bounded Counts}

\begin{lemma}\label{lem:counts}
  $\Pr[\overline{\EvT_L}]\le L\,e^{\Lambda}\,e^{-2R/3}\le\varphi/3$.
\end{lemma}

\begin{proof}
  We have $\Pr[\overline{\EvT_L}]\le\sum_{i=1}^L\Pr[\EvT_{i-1}\cap\overline{\Ev_i}]$.
  Fix $h\in\Hset$ and $i$, and condition on $\Fam_{i-1}$.
  The $m$ indicators $[Q_i[u]\in h]$ are independent, and by \cref{lem:radon} each is $1$ only if two of the $t$ chosen indices of $Q_{i-1}$ lie in $h$, which by a union bound over the $c$ pairs of positions has probability at most $c\,\ff{r}{2}/\ff{m'}{2}\le c\,(r/m')^2$, where $r=r_h(Q_{i-1})$ and $m'=|Q_{i-1}|$.
  We use $c\ge6$, $R\ge1$, and $8ecR\le m\le 8ecR+1$.

  For $i=1$, $r\le k_0-1<\beta\nu$ and $m'=\nu$, so the probability is at most $c\beta^2=\frac{1}{64c}$ and $\E[r_h(Q_1)]\le\frac{m}{64c}\le\frac{eR}{8}+ \frac{1}{64c}$.
  By \cref{lem:chernoff} with $a=2R$, $\Pr[r_h(Q_1)\ge 2R]\le\bigl(\frac{e^2}{16}+\frac{e}{128cR}\bigr)^{2R}\le0.47^{2R}\le e^{-2R/3}$.

  For $i\ge2$, on $\EvT_{i-1}$ we have $r\le2R$ and $m'=m$, so $\E[r_h(Q_i)\mid\Fam_{i-1}]\le 4cR^2/m\le R/(2e)$, and \cref{lem:chernoff} with $a=R$ gives $\Pr[r_h(Q_i)\ge R\mid\Fam_{i-1}]\le2^{-R}\le e^{-2R/3}$.

  A union bound over the at most $e^{\Lambda}$ halfspaces $h\in\Hset$ and over $i\in[L]$ gives the first inequality, and the second follows from $R\ge\frac32(\Lambda+\ln\frac{3L}\varphi)$.
\end{proof}

\subsection{The First Layer}

\begin{lemma}\label{lem:first}
  For all $k\ge1$, $A_1(k)\le\min\bigl\{1,\;e^{\Lambda}\,(c\beta^2)^k\bigr\} =\min\bigl\{1,\;e^{\mu_1(\kappa_1-k)}\bigr\}$, where $\mu_1:=\ln(64c)$ and $\kappa_1=\Lambda/\mu_1$.
\end{lemma}

\begin{proof}
  By \cref{obs:pi}, $A_1(k)\le\E[\pi_k(Q_1)]\le\sum_{h}\E[\ff{r_h(Q_1)}{k}]/\ff{m}{k}$.
  As in the proof of \cref{lem:counts}, $r_h(Q_1)$ is a sum of $m$ independent indicators, each with probability at most $p:=c\beta^2$.
  The falling factorial $\ff{r_h(Q_1)}{k}$ counts the ordered $k$-tuples of distinct indices whose indicators are all $1$, so its expectation is at most $\ff{m}{k}p^k$.
  Summing over the at most $e^\Lambda$ halfspaces gives $A_1(k)\le e^{\Lambda}p^k=e^{\Lambda-\mu_1k}$, since $1/p=64c$.
\end{proof}

\subsection{The Main Phase: Shrinking the Threshold}

The next lemma is the heart of the analysis.
It applies to any layer $i\ge1$ whose tuple probabilities decay exponentially, and it shows that the next layer inherits an exponential bound with a smaller ``threshold.''

\begin{lemma}[Step]\label{lem:step}
  Let $1\le i<L$, $\Lambda'\ge0$, and $\mu>0$, and suppose that $A_i(j)\le\min\{1,e^{\Lambda'-\mu j}\}$ for all $j\ge1$.
  Let $1\le k\le m/2$ and $\ell_k:=\ln\frac{m}{2k}$.
  Then, with $\theta:=\min\{1,\ell_k/\mu\}$,
  \[
    A_{i+1}(k)\le\tfrac14\exp\bigl(\theta\Lambda'-k\,(2\theta\mu-\ln(4c))\bigr).
  \]
\end{lemma}

\begin{proof}
  Since $\min\{1,e^y\}\le e^{\theta y}$ for $\theta\in[0,1]$, \cref{lem:recursion} gives $A_{i+1}(k)\le c^ke^{\theta\Lambda'}\,\E[w^{|U_k|}]$ with $w:=e^{-\theta\mu}\in(0,1]$ and $|Q_i|=m$.
  Since $\theta\mu\le\ell_k$ we have $w\ge 2k/m$, and \cref{lem:collisions} yields $\E[w^{|U_k|}]\le w^2(2w)^{2k-2}=\frac14(4w^2)^k$.
  Hence $A_{i+1}(k)\le\frac14 e^{\theta\Lambda'}(4cw^2)^k$, which is the claim.
\end{proof}

\begin{lemma}[Main phase]\label{lem:main-phase}
  Let $1\le i<L$, $\kappa\ge0$, and $\mu\ge\ell$, and suppose that $A_i(j)\le\min\{1, e^{\mu(\kappa-j)}\}$ for all $j\ge1$.
  Let $\mu':=2\ell-\ln(4c)$ and $\kappa':=\ell\kappa/\mu'$.
  Then $A_{i+1}(k)\le\min\{1,e^{\mu'(\kappa'-k)}\}$ for all $k\ge1$.
  Moreover, $\mu'\ge\ell+1$ and $\kappa'\le\psi\kappa$, where $\psi:=\frac{\ln(4c)+1}{\ln(4c)+2}$.
\end{lemma}

\begin{proof}
  Since $i+1\ge2$, we have $A_{i+1}(k)=0$ for $k>R$.
  Let $1\le k\le R$; then $k\le m/2$ and $\ell_k=\ln\frac{m}{2k}\ge\ell$.
  We apply \cref{lem:step} with $\Lambda'=\mu\kappa$.
  The exponent becomes
  \[
    \theta\mu\kappa-2k\theta\mu+k\ln(4c)=\theta\mu\,(\kappa-2k)+k\ln(4c),
  \]
  where $\theta\mu=\min\{\mu,\ell_k\}\ge\ell$.
  If $2k\ge\kappa$, the exponent is at most $\ell(\kappa-2k)+k\ln(4c)=\ell\kappa-k\mu'=\mu'(\kappa'-k)$, as claimed.
  If $2k<\kappa$, then $\mu'(\kappa'-k)=\ell\kappa-\mu'k>\ell\kappa-2\ell k>0$ (using $\mu'<2\ell$), so the claimed bound is $1$ and holds trivially.
  Finally, $\mu'-\ell=\ell-\ln(4c)\ge1$, and $\kappa'/\kappa=\ell/(2\ell-\ln(4c))$ is decreasing in $\ell$, so $\ell\ge\ln(4c)+1$ gives $\kappa'\le\psi\kappa$.
\end{proof}

\begin{corollary}\label{cor:main-phase}
  Let $i^*:=1+L_1$.
  There are $\kappa^*\le1$ and $\mu^*\ge\ell$ such that $A_{i^*}(j)\le\min\{1,e^{\mu^*(\kappa^*-j)}\}$ for all $j\ge1$.
\end{corollary}

\begin{proof}
  By \cref{lem:first}, the hypothesis of \cref{lem:main-phase} holds for $i=1$ with $(\kappa_1,\mu_1)$, and $\mu_1=\ln(64c)\ge\ln(4ec+\frac12)\ge\ell$ since $m\le8ecR+1$.
  By induction, after $L_1$ applications of \cref{lem:main-phase} we have $\kappa^*\le\psi^{L_1}\kappa_1$ and $\mu^*\ge\ell$.
  Since $\ln\frac1\psi=\ln\bigl(1+\frac{1}{\ln(4c)+1}\bigr)\ge\frac{1}{\ln(4c)+2}$, the choice of $L_1$ gives $\psi^{L_1}\kappa_1\le1$ (if $\kappa_1\le1$ then $L_1=0$).
\end{proof}

\subsection{The Final Phase}

Once the threshold is below $1$, the bound $A_i(j)\le e^{\mu^*\kappa^*}(e^{-\mu^*})^j$ is a prefactor times a geometric term, and by the next lemma the prefactor shrinks by a constant factor per layer.

\begin{lemma}[Final phase]\label{lem:final-phase}
  Let $1\le i<L$, $\eta>0$, and $0<a\le x$, and suppose that $A_i(j)\le\eta a^j$ for all $j\ge2$.
  Then $A_{i+1}(k)\le\eta'(a')^k$ for all $k\ge2$, where $\eta'=\eta a^2/(4x^2)$ and $a'=4cx^2\le x/e$.
\end{lemma}

\begin{proof}
  Since $i+1\ge2$, we have $A_{i+1}(k)=0$ for $k>R$.
  For $2\le k\le R$, since $|U_k|\ge2$, \cref{lem:recursion,lem:collisions} give $A_{i+1}(k)\le c^k\eta\,\E[a^{|U_k|}]\le c^k\eta a^2(a+2k/m)^{2k-2}$.
  Since $a\le x$ and $2k/m\le x$, this is at most $c^k\eta a^2(2x)^{2k-2}=\frac{\eta a^2}{4x^2}(4cx^2)^k$.
  Finally, $4cx\le\frac1e$ since $x\le\frac1{4ec}$.
\end{proof}

\subsection{Putting Things Together}

\begin{theorem}\label{thm:main}
  Let $P$ be a set of $n$ points in general position in $\R^d$, $d\ge2$, and let $\eps,\varphi\in(0,1)$.
  With probability at least $1-\varphi$, \cref{alg:layered} returns a $(1-\eps)\beta$-centerpoint of $P$, where $\beta=\frac{1}{4(d+1)(d+2)}$.
  It computes at most $Lm+1$ Radon points, where
  \[
    m=O\bigl(d^2\,(d\log\tfrac d\eps+\log\tfrac1\varphi)\bigr)
    \quad\text{and}\quad
    L=O\bigl(\log d\,\bigl(\log d+\lnp\log\tfrac1\eps+\lnp\log\tfrac1\varphi\bigr)+\log\tfrac1\varphi\bigr),
  \]
  and its running time is $O\bigl(n_0(d+\log n_0)\log\frac6\varphi+d^{\omega}Lm\bigr)$ with $n_0=O(\eps^{-2}d^2(d\log d+\log\frac1\varphi))$.
\end{theorem}

\begin{proof}
  \emph{Correctness.}
  By \cref{cor:main-phase}, at layer $i^*=1+L_1$ we have $A_{i^*}(j)\le\eta a^j$ for $j\ge1$ with $\eta=e^{\mu^*\kappa^*}\le e^{\mu^*}$ and $a=e^{-\mu^*}\le e^{-\ell}=x$.
  Applying \cref{lem:final-phase} once gives, at layer $i^*+1$, $\eta_{i^*+1}=\eta a^2/(4x^2)\le e^{-\mu^*}/(4x^2)\le 1/(4x)$ and $a_{i^*+1}=4cx^2\le x$.
  Each further application multiplies the prefactor by $a'^2/(4x^2)=4c^2x^2\le\frac{1}{4e^2}$ and keeps $a=4cx^2$.
  Hence, at the last layer $L=i^*+L_2$,
  \[
    c\,A_L(2)\le c\cdot\frac{1}{4x}\,(4c^2x^2)^{L_2-1}\,(4cx^2)^2
    =4c^3x^3\,(4c^2x^2)^{L_2-1}\le\frac{1}{16e^3}\,(4e^2)^{-(L_2-1)}\le\frac{\varphi}{3},
  \]
  where we used $cx\le\frac1{4e}$ and $(4e^2)^{-(L_2-1)}\le e^{-3(L_2-1)}\le\varphi$.
  By \cref{lem:output,lem:counts}, $z$ lies in a bad halfspace with probability at most $2\varphi/3$.
  Otherwise, by \cref{lem:bad-family,lem:sample-ok}, $z$ is a $k_0/\nu$-centerpoint of $S$, and $k_0/\nu\ge\beta$.
  By \cref{lem:transfer} with $\rho=\beta/2\le\beta/(1+\eps)$, $z$ is a $(1-\eps)\beta$-centerpoint of $P$.
  Adding the failure probability $\varphi/3$ of \cref{lem:sample-ok} proves correctness.

  \emph{Complexity.}
  The algorithm computes $Lm+1$ Radon points, each in $O(d^\omega)$ time by \cref{lem:radon}; choosing a random sequence of distinct indices takes $O(d)$ expected time per Radon point (rejection), or $O(d\log d)$ time deterministically.
  Each sampling attempt takes $O(n_0(d+\log n_0))$ time (drawing the points and sorting their indices to detect repetitions), and there are at most $\lceil\log_2(6/\varphi)\rceil$ attempts.
  For the bounds on $m$ and $L$, note that $1/\beta=O(d^2)$, so $n_0=O(\eps^{-2}d^2(d\log d+\log\frac1\varphi))$ by \eqref{eq:n0}, and $\nu\le n_0^2$ (if $S=P$ then $n\le n_0^2$).
  Hence $\Lambda=\ln2+d\ln\frac{e\nu}{d}=O(d\log n_0)=O(d\log\frac d\eps+d\,\lnp\log\frac1\varphi) =O(d\log\frac d\eps+\log\frac1\varphi)$, where the last step uses $d\ln y\le d\ln d+y$ for $y>0$.
  Moreover $\kappa_1=O(\Lambda/\log d)$, so $\lnp\kappa_1=O(\log d+\lnp\log\frac1\eps+\lnp\log\frac1\varphi)$, $L_1=O(\log d\cdot\lnp\kappa_1)$, and $L_2=O(1+\log\frac1\varphi)$, which gives the bound on $L$.
  Finally $R=O(\Lambda+\log\frac L\varphi)$ and $m=O(cR)$.
\end{proof}

For constant $\eps$ and $\varphi$ we have $m=O(d^3\log d)$ and $L=O(\log^2 d)$, which gives \cref{thm:intro}: $O(d^3\log^3 d)$ Radon points and $O(d^{3+\omega}\log^3d)$ time, or $O(d^6\log^3 d)$ time with Gaussian elimination.
For general $\eps$ and $\varphi$, the number of layers grows by an additive $O(\log\frac1\varphi+\log d\,\lnp\log\frac1\eps)$, since $\log d\cdot\lnp\log\frac1\varphi\le\log^2d+\log\frac1\varphi$.

\section{Layers of Tverberg Points}\label{sec:tverberg}

An \emph{$s$-Tverberg partition} of a sequence $X$ of points is a partition of its index set into $s$ nonempty parts whose convex hulls have a common point, called an \emph{$s$-Tverberg point} of~$X$.
A Radon point is a $2$-Tverberg point.
In this section we replace the Radon points of \cref{alg:layered} by $s$-Tverberg points with $s=\Theta(\log d)$, computed by linear programming as in~\cite{dp-ncptad-26}.
Throughout, $s\ge3$ is an integer, and the number $t$ of inputs of a gate replaces $d+2$:
\[
  D:=(s-1)(d+1),\qquad t:=\bigl\lceil 3sD\ln(3s)\bigr\rceil,\qquad c_s:=\binom ts .
\]
We assume $n\ge t$; then the gates always have $t$ distinct inputs (see the proof of \cref{thm:tverberg}).

\subsection{Tverberg Gates}

The next lemma is Lemma~1.3 of~\cite{dp-ncptad-26} with slightly different constants; it is based on Sarkaria's lifting~\cite{s-ttnf-92} and Sober\'on's random partitions~\cite{s-tpwen-19}.

\begin{lemma}\label{lem:tverberg}
  Let $X=(x_1,\dots,x_t)$ be a sequence of points of $\R^d$, and let $\chi:[t]\to[s]$ be a uniformly random coloring.
  With probability at least $\frac12$, the color classes form an $s$-Tverberg partition of $X$.
  In that case, a common point $p$ of their convex hulls is obtained from any feasible solution of a linear feasibility problem with $t$ nonnegative variables and $D+1$ equality constraints, whose coefficients are $0$, $\pm1$, and the coordinates of the points of~$X$ and their negatives.
  Every halfspace (open or closed) that contains $p$ contains points of $X$ at $s$ distinct indices, one of each color.
\end{lemma}

\begin{proof}[Proof of \cref{lem:tverberg}]
  Let $v_1,\dots,v_{s-1}$ be the standard basis of $\R^{s-1}$ and $v_s:=-\sum_{j<s}v_j$.
  Then $\sum_j\alpha_jv_j=0$ holds if and only if $\alpha_1=\dots=\alpha_s$.
  For $i\in[t]$ and $j\in[s]$, let $y_{i,j}:=(x_i,1)\otimes v_j\in\R^{d+1}\otimes\R^{s-1}\cong\R^D$, and let $y_i:=y_{i,\chi(i)}$.
  Consider the problem
  \[
    \lambda\in\R^t,\quad\lambda\ge0,\quad\textstyle\sum_i\lambda_i=1,\quad\sum_i\lambda_iy_i=0 .
  \]
  Suppose that $\lambda$ is feasible, and let $w_j:=\sum_{i:\chi(i)=j}\lambda_i(x_i,1)$.
  Then $\sum_jw_j\otimes v_j=0$.
  Looking at each of the $d+1$ coordinates of the $w_j$, the property of the $v_j$ gives $w_1=\dots=w_s$.
  The last coordinate shows that every color class has total weight $\sum_{i:\chi(i)=j}\lambda_i=1/s$.
  Hence $p:=s\sum_{i:\chi(i)=j}\lambda_ix_i$ does not depend on $j$, and it is a convex combination of the points of color $j$ for every $j$.
  If a halfspace $h$ contains no point of color $j$, then the convex complement of $h$ contains all points of color $j$ and hence $p$.

  The problem is infeasible if and only if $0\notin\conv\{y_1,\dots,y_t\}$, that is, if and only if there is $u\in\R^D$ with $\langle u,y_i\rangle<0$ for all $i$.
  Fix $u$.
  Since $\sum_jy_{i,j}=(x_i,1)\otimes\sum_jv_j=0$, some $j$ has $\langle u,y_{i,j}\rangle\ge0$.
  So $\langle u,y_i\rangle<0$ with probability at most $1-\frac1s$, independently for all $i$, and the probability that this happens for all $i$ is at most $e^{-t/s}$.

  This event depends only on the set of vectors $y_{i,j}$ with $\langle u,y_{i,j}\rangle\ge0$.
  Closed halfspaces through the origin in $\R^D$ have VC-dimension at most~$D$.
  Indeed, among any $D+1$ vectors there is a nontrivial linear relation $\sum_q\alpha_qz_q=0$; we may assume that $I:=\{q:\alpha_q>0\}\ne\emptyset$.
  No $u$ has $\langle u,z_q\rangle<0$ exactly for $q\in I$, since then $\sum_q\alpha_q\langle u,z_q\rangle<0$.
  By the Sauer--Shelah lemma~\cite{vc-ucrfe-71,s-dfs-72,s-cpsom-72,m-ldg-02}, $u$ induces at most $(est/D)^D$ distinct sets on the $st$ vectors $y_{i,j}$.
  A union bound therefore bounds the failure probability by $(est/D)^De^{-t/s}=\exp\bigl(D(\ln(e\zeta s^2)-\zeta)\bigr)$, where $\zeta:=t/(sD)\ge3\ln(3s)$.
  The function $\zeta\mapsto\ln(e\zeta s^2)-\zeta$ is decreasing for $\zeta\ge1$.
  At $\zeta=3\ln(3s)$ it equals $1-2\ln3-(\ln s-\ln\ln(3s))\le1-2\ln3-(\ln3-\ln\ln9)<-1.5$, since $\ln s-\ln\ln(3s)$ is increasing in $s$.
  Hence the failure probability is at most $e^{-1.5}<\frac12$.
\end{proof}

A \emph{Tverberg gate} applied to $X$ draws random colorings until the problem of \cref{lem:tverberg} is feasible, solves it, and returns~$p$.
It solves at most two linear programs in expectation.
The analysis below uses only the last property of \cref{lem:tverberg}, which holds for every successful attempt, so unlike in~\cite{dp-ncptad-26} no failure event is needed.

\subsection{Algorithm and Analysis}

\textsc{LayeredTverberg} is \cref{alg:layered} with two changes: every Radon computation, including the output, is replaced by a Tverberg gate applied to $t$ points at distinct random indices, and the parameters are adapted to the gates.
The quality is $\beta:=(x_0/(2e^2s\,c_s))^{1/s}$, where $x_0:=(2^sc_s)^{-1/(s-2)}$ satisfies $2^sc_sx_0^{s-2}=1$; the sample $S$, $\nu$, $k_0$, and $\Lambda$ are defined as in \cref{sec:algorithm} with this $\beta$.
The number of layers is $L=1+L_1+L_2$ with $L_1=\lceil\log_2^+\kappa_1\rceil$, where $\kappa_1:=\Lambda/\ln\frac{1}{c_s\beta^s}$ and $\log_2^+y=\max\{0,\log_2y\}$, and $L_2=\max\{1,\lceil\frac1s\ln\frac3\varphi\rceil\}$.
The count threshold is $R=\lceil\Lambda+\ln\frac{3L}{\varphi}\rceil$ for all layers, and the layers have size $m=\lceil sR/x_0\rceil$; let $x:=sR/m\le x_0$ and $\ell:=\ln\frac1x$.

We keep the notation of \cref{sec:algorithm}: $\pi_k$, $r_h$, $\Ev_i$, $\EvT_i$, and $A_i(k)$, where now $\Ev_i$ is the event that $r_h(Q_i)\le R$ for all $h\in\Hset$, for every $i\ge1$.
Let $U^{(s)}_k$ be the union of $k$ independent uniformly random $s$-subsets of $[m']$.
We have
\begin{equation}\label{eq:tv-ell}
  (s-2)\,\ell\;\ge\;(s-2)\ln\tfrac1{x_0}\;=\;\ln(2^sc_s).
\end{equation}

The next lemma collects the analogues of \cref{lem:output,lem:recursion,lem:collisions,lem:counts,lem:first}.

\begin{lemma}\label{lem:tv-basic}
  Assume the outcome of \cref{lem:sample-ok}.
  Then:
  \begin{enumerate}
    \item $\Pr[z\in\bigcup\Hset]\le\Pr[\overline{\EvT_L}]+c_s\,A_L(s)$;
    \item $A_{i+1}(k)\le\min\{1,c_s^k\sum_j\Pr[|U^{(s)}_k|=j]\,A_i(j)\}$ for $1\le i<L$, $k\ge1$, and $m'=m$;
    \item $\E[w^{|U^{(s)}_k|}]\le w^s(w+sk/m')^{s(k-1)}$ for $m'\ge s$, $k\ge1$, and $w\in(0,1]$;
    \item $A_1(k)\le\min\{1,e^\Lambda(c_s\beta^s)^k\}$ for all $k\ge1$;
    \item $\Pr[\overline{\EvT_L}]\le Le^\Lambda e^{-R}\le\varphi/3$.
  \end{enumerate}
\end{lemma}

\begin{proof}[Proof of \cref{lem:tv-basic}]
  Items 1, 2, and 4 are proved like \cref{lem:output,lem:recursion,lem:first}.
  The last property of \cref{lem:tverberg} replaces \cref{lem:radon}, the $c_s$ choices of $s$ positions among $t$ replace the $c$ pairs, and $\ff{r}{s}/\ff{m'}{s}\le(r/m')^s$ holds for $r\le m'$.

  For item 3, reveal the $sk$ indices one at a time, subset by subset.
  An index that is the $(g+1)$-st element of its subset, $0\le g<s$, is uniform over the $m'-g$ indices not yet used by its subset.
  If $\Delta$ distinct indices have been revealed so far, it repeats one of them with probability $(\Delta-g)/(m'-g)\le\Delta/m'\le sk/m'$.
  The first $s$ indices are distinct, and multiplying the bounds $\E[w^{\xi}\mid\cdot]\le w+sk/m'$ for the other $s(k-1)$ indices gives the claim, as in \cref{lem:collisions}.

  For item 5, fix $h\in\Hset$ and $i$, and condition on $\Fam_{i-1}$.
  The indicators $[Q_i[u]\in h]$ are independent, and each has probability at most $c_s(r/m')^s$ with $r=r_h(Q_{i-1})$.
  For $i=1$ this is at most $c_s\beta^s=x_0/(2e^2s)$, so $\E[r_h(Q_1)]\le(sR/x_0+1)\,x_0/(2e^2s)\le R/e^2$.
  For $i\ge2$, on $\EvT_{i-1}$ it is at most $c_s(R/m)^s=c_s(x/s)^s$, and since $m=sR/x$ and $c_sx^{s-1}\le c_sx_0^{s-1}=2^{-s}x_0$, we get $\E[r_h(Q_i)\mid\Fam_{i-1}]\le Rc_sx^{s-1}s^{1-s}\le R/e^2$.
  In both cases \cref{lem:chernoff} with $a=R$ gives probability at most $e^{-R}$.
  The union bound over $h$ and $i$ and the choice of $R$ complete the proof.
\end{proof}

\begin{lemma}[Main phase with Tverberg gates]\label{lem:tv-main}
  Let $1\le i<L$, $\kappa\ge0$, and $\mu\ge\ell$, and suppose that $A_i(j)\le\min\{1,e^{\mu(\kappa-j)}\}$ for all $j\ge1$.
  Let $\mu':=s\ell-\ln(2^sc_s)$ and $\kappa':=\ell\kappa/\mu'$.
  Then $A_{i+1}(k)\le\min\{1,e^{\mu'(\kappa'-k)}\}$ for all $k\ge1$, $\mu'\ge2\ell$, and $\kappa'\le\kappa/2$.
\end{lemma}

\begin{proof}
  The inequality $\mu'\ge2\ell$ is~\eqref{eq:tv-ell}, and it implies $\kappa'\le\kappa/2$.
  We have $A_{i+1}(k)=0$ for $k>R$.
  Let $1\le k\le R$, $\ell_k:=\ln\frac{m}{sk}\ge\ell$, $\theta:=\min\{1,\ell_k/\mu\}$, and $w:=e^{-\theta\mu}\ge sk/m$.
  As in \cref{lem:step}, items 2 and 3 of \cref{lem:tv-basic} give
  \[
    A_{i+1}(k)\le c_s^ke^{\theta\mu\kappa}w^s(2w)^{s(k-1)}
    =2^{-s}\exp\bigl(\theta\mu(\kappa-sk)+k\ln(2^sc_s)\bigr),
  \]
  and $\theta\mu=\min\{\mu,\ell_k\}\ge\ell$.
  If $sk\ge\kappa$, the exponent is at most $\ell(\kappa-sk)+k\ln(2^sc_s)=\mu'(\kappa'-k)$.
  If $sk<\kappa$, then $\mu'(\kappa'-k)\ge\ell(\kappa-sk)>0$ since $\mu'\le s\ell$, so the bound is~$1$.
\end{proof}

For $s=2$ (\cref{lem:main-phase}), $\mu'-\ell=\ell-\ln(4c)$ is only a constant for layers of size $\Theta(cR)$, so the threshold shrinks only by a factor $1-\Theta(1/\log d)$ per layer.
For $s\ge3$, the extra factors $w^{s-2}$ pay for the term $k\ln(2^sc_s)$, and the threshold halves in every layer.

\begin{lemma}[Final phase with Tverberg gates]\label{lem:tv-final}
  Let $1\le i<L$, $\eta>0$, and $0<a\le x$, and suppose that $A_i(j)\le\eta a^j$ for all $j\ge s$.
  Then $A_{i+1}(k)\le\eta(a/(2x))^s\,(a')^k$ for all $k\ge1$, where $a':=c_s(2x)^s\le x^2$.
\end{lemma}

\begin{proof}[Proof of \cref{lem:tv-final}]
  For $k>R$ the claim holds since $A_{i+1}(k)=0$.
  For $k\le R$ we have $|U^{(s)}_k|\ge s$ and $sk/m\le x$, so items 2 and 3 of \cref{lem:tv-basic} give $A_{i+1}(k)\le c_s^k\eta a^s(2x)^{s(k-1)}$, which is the claim.
  Finally, $c_s(2x)^s=2^sc_sx^{s-2}\,x^2\le2^sc_sx_0^{s-2}x^2=x^2$.
\end{proof}

\begin{theorem}\label{thm:tverberg}
  Let $P$ be a set of $n$ points in general position in $\R^d$, $d\ge2$, let $\eps,\varphi\in(0,1)$ and $3\le s\le2d$, and assume $n\ge t$.
  With probability at least $1-\varphi$, \textup{\textsc{LayeredTverberg}} returns a $(1-\eps)\beta$-centerpoint of $P$, where $\beta$ is defined above.
  It applies $Lm+1$ Tverberg gates, and hence solves at most $2(Lm+1)$ linear programs in expectation, each with $t$ variables and $D+1$ constraints.
  For $s=\max\{3,\lceil\ln d\rceil\}$ we have
  \[
    \beta=\Theta\Bigl(\frac{1}{d\log d\log\log d}\Bigr),\quad
    t=\Theta(d\log^2d\log\log d),\quad
    m=O\bigl(t\,(d\log\tfrac d\eps+\log\tfrac1\varphi)\bigr),
  \]
  and $L=O(\log d+\lnp\log\frac1\eps+\log\frac1\varphi)$.
\end{theorem}

The proof follows that of \cref{thm:main}.
By item~4 of \cref{lem:tv-basic} and $L_1$ applications of \cref{lem:tv-main}, the threshold drops below~$1$ at layer $1+L_1$.
Then $L_2$ applications of \cref{lem:tv-final} give $c_sA_L(s)\le x^{sL_2}\le\varphi/3$, and items~1 and~5 of \cref{lem:tv-basic} bound the failure probability.

For constant $\eps$ and $\varphi$, \cref{thm:tverberg} applies $Lm+1=O(d^2\log^4d\log\log d)$ Tverberg gates, which proves \cref{thm:intro-tverberg}.
Setting up a linear program takes only $O(tD)$ time.
The detailed proof follows below.

\begin{proof}[Proof of \cref{thm:tverberg}]
  \emph{Correctness.}
  We use $c_s\ge(t/s)^s$, which gives $x_0\le(2t/s)^{-s/(s-2)}\le s/(2t)<1/e$.
  Since $m\le sR/x_0+1\le2sR/x_0$, we have $x\ge x_0/2$ and hence $\mu_1:=\ln\frac1{c_s\beta^s}=\ln\frac{2e^2s}{x_0}\ge\ln\frac1x=\ell$.
  By item 4 of \cref{lem:tv-basic} and $L_1$ applications of \cref{lem:tv-main}, at layer $i^*=1+L_1$ we have $A_{i^*}(j)\le\min\{1,e^{\mu^*(\kappa^*-j)}\}$ with $\kappa^*\le1$ and $\mu^*\ge\ell$.
  Hence $A_{i^*}(j)\le\eta a^j$ with $\eta=e^{\mu^*\kappa^*}\le e^{\mu^*}$ and $a=e^{-\mu^*}\le x$.
  One application of \cref{lem:tv-final} gives the prefactor $e^{\mu^*}e^{-s\mu^*}(2x)^{-s}\le x^{s-1}(2x)^{-s}=2^{-s}/x$ and $a'\le x^2$.
  Each further application multiplies the prefactor by $(a'/(2x))^s\le x^s$.
  At layer $L=i^*+L_2$,
  \[
    c_sA_L(s)\le c_s\,\frac{2^{-s}}{x}\,x^{s(L_2-1)}\,x^{2s}
    \le\bigl(c_sx^{s-2}\bigr)\,x^{s+1}\,x^{s(L_2-1)}\le x^{sL_2}\le e^{-sL_2}\le\frac\varphi3,
  \]
  using $2^{-s}\le1$, $c_sx^{s-2}\le2^{-s}$, and $x<1/e$.
  The rest is as in the proof of \cref{thm:main}: \cref{lem:sample-ok} holds with the same proof, since $\nu>1/\beta\ge t/s\ge2d+2$ by $\beta\le s/t$ (see below).
  Moreover, $\nu\ge t$: if $S=P$ then $\nu=n\ge t$, and otherwise $\nu=n_0\ge\frac{2d}{\beta}\ge\frac{2dt}{s}\ge t$ by $s\le2d$.
  Also $m\ge sR/x_0\ge2tR\ge t$ since $x_0\le s/(2t)$, so the gates always have $t$ distinct inputs.
  The quality is $k_0/\nu\ge\beta$, and \cref{lem:transfer} applies with $\rho=\beta/2$.

  \emph{Parameters.}
  From $(t/s)^s\le c_s\le(et/s)^s$ and the definition of $\beta$,
  \[
    \frac{s}{et}\Bigl(\frac{2et}{s}\Bigr)^{-1/(s-2)}(2e^2s)^{-1/s}\le\beta\le\frac st .
  \]
  For $s=\max\{3,\lceil\ln d\rceil\}$ we have $\ln\frac{2et}{s}=O(s)$, so for $d$ large enough that $s=\lceil\ln d\rceil$, both correction factors are bounded below by constants.
  Hence $\beta=\Theta(s/t)$, and $t=\Theta(s^2d\log s)$ gives the bounds on $\beta$ and~$t$.
  Similarly, $1/x_0\le(2et/s)^{s/(s-2)}=O(t/s)$, so $m=O(tR)$.
  As in the proof of \cref{thm:main}, $n_0=O(\eps^{-2}\beta^{-1}(d\log\frac1\beta+\log\frac1\varphi))$ gives $\Lambda=O(d\log\frac d\eps+\log\frac1\varphi)$ and $R=O(\Lambda+\log\frac L\varphi)$.
  Finally, $L_1=O(\log\Lambda)$ and $L_2=O(1+\log\frac1\varphi)$.
\end{proof}

\section{Bit Length}\label{sec:bits}

In the real RAM, \cref{alg:layered} performs $O(d^\omega)$ arithmetic operations per Radon point, independently of the input.
On a Turing machine, exact computation is expensive.
The Radon point of points with $b$-bit coordinates may need up to $O(db)$ bits, so the bit length can grow by a factor $\Theta(d)$ per layer, and after $L=\Theta(\log^2 d)$ layers it may be $d^{\Theta(\log^2 d)}$ times the input bit length.
Dutta and Pisula~\cite[Theorem~1.2]{dp-ncptad-26} address the same issue for their algorithm: truncating every computed point to $O(d^3\log^4 d\log\frac d{p_f}+\log\frac1{\eps'})$ bits, where $p_f$ is the failure probability, keeps the output within distance $\eps'$ of the output of the exact computation.
We show that rounding every computed point to a fine grid affects the analysis only through a loss of $d$ points per bad halfspace, so the rounded computation itself outputs a centerpoint.
The argument uses only the property of \cref{lem:radon}, so it also applies to the Tverberg points of \cref{sec:tverberg}.

Assume that the points of $P$ have integer coordinates of absolute value at most $2^B$.
For an integer $b\ge1$, let $\tau:=\sqrt d\,2^{-b}$, and let \emph{rounding} map a point to a nearest point of the grid $2^{-b}\mathbb{Z}^d$, which moves it by at most $\tau$.
For an open halfspace $h=\{y:\langle a,y\rangle>\alpha\}$ with $\|a\|=1$ and for $\lambda\ge0$, let $h^{+\lambda}:=\{y:\langle a,y\rangle>\alpha-\lambda\}$.
If $y\in h^{+\lambda}$ and $\|y-y'\|\le\tau$, then $y'\in h^{+\lambda+\tau}$.

\begin{lemma}\label{lem:gap}
  Let $S\subset\mathbb{Z}^d$ be a set of at least $d+1$ points in general position with coordinates of absolute value at most $2^B$.
  For every hyperplane $H$ spanned by $d$ points of $S$ and every $y\in S\setminus H$, the distance from $y$ to $H$ is at least $\delta:=d^{-d/2}\,2^{-(B+1)(d-1)}$.
\end{lemma}

\begin{proof}
  Let $H$ be spanned by $y_1,\dots,y_d\in S$; these points are affinely independent, since otherwise they and any other point of $S$ would lie on a common hyperplane.
  Let $a\in\mathbb{Z}^d$ be the vector of signed $(d-1)\times(d-1)$ minors of the $(d-1)\times d$ matrix with rows $y_j-y_1$, $2\le j\le d$.
  Then $a\ne0$, $a$ is orthogonal to $H-y_1$, and by Hadamard's inequality every entry of $a$ has absolute value at most $(\sqrt{d-1}\,2^{B+1})^{d-1}$, so $\|a\|\le d^{d/2}2^{(B+1)(d-1)}$.
  For $y\in S\setminus H$, $\langle a,y-y_1\rangle$ is a nonzero integer, and the distance is $|\langle a,y-y_1\rangle|/\|a\|\ge1/\|a\|$.
\end{proof}

\begin{proposition}\label{prop:bits}
  Let $P\subset\mathbb{Z}^d$ be a set of $n\ge n_0$ points in general position with coordinates of absolute value at most $2^B$, where $n_0$ is as in \cref{alg:layered}, and let $\eps\le\frac13$.
  Modify \cref{alg:layered} so that every point of $Q_1,\dots,Q_L$ is rounded to $2^{-b}\mathbb{Z}^d$ right after it is computed, where $b$ is the smallest integer with $L\tau<\delta$.
  Then $b=O(d(B+\log d)+\log L)$, and with probability at least $1-\varphi$ the output is a $(1-\eps)^2\beta$-centerpoint of~$P$.
  Every Radon point is computed from points whose coordinates are integer multiples of $2^{-b}$ of absolute value at most $2^B+1$.
\end{proposition}

\begin{proof}
  The bound on $b$ is immediate from \cref{lem:gap}.
  Every Radon point lies in the convex hull of its inputs and rounding moves each coordinate by at most $2^{-b-1}$, so by induction all coordinates of layer $i$ have absolute value at most $2^B+i2^{-b-1}\le2^B+1$.

  Since $n\ge n_0$, the sample has $\nu\ge n_0\ge\frac{2d}{\eps^2\beta}\ge\frac{d}{\eps\beta}$ points.
  Let $k_0':=\lceil\beta\nu\rceil-d$, so that $1\le k_0'\le\nu-d$, and let $\Hset'$ be the family of \cref{lem:bad-family} for $S$ and $k_0'$.
  For $h\in\Hset'$, the set $h^{+L\tau}$ contains no point of $S$ outside the closure of $h$, by \cref{lem:gap} and $L\tau<\delta$.
  The closure of $h$ contains at most $k_0'-1$ points of $S$ in $h$ and $d$ on its boundary.
  Hence
  \begin{equation}\label{eq:robust-count}
    |h^{+L\tau}\cap S|\le\lceil\beta\nu\rceil-1<\beta\nu .
  \end{equation}
  We repeat the analysis of \cref{sec:algorithm} with $\Hset'$ in place of $\Hset$, measuring layer $i$ against the enlarged halfspaces $h^{+(L-i)\tau}$.
  That is, $\pi_k(Q_i)$ is the probability that $Q_i[J]\subseteq h^{+(L-i)\tau}$ for some $h\in\Hset'$, and $r_h(Q_i)$ counts the indices of $Q_i$ in $h^{+(L-i)\tau}$.
  The analysis uses three facts about $\Hset$ and the Radon points:
  \begin{itemize}
    \item $|\Hset|\le e^\Lambda$;
    \item every bad halfspace contains fewer than $\beta\nu$ points of $S=Q_0$;
    \item if a point of layer $i+1$ (or the output, for $i=L$) lies in a halfspace of the family for layer $i+1$ (or in $h$ itself), then the halfspace of the family for layer $i$ contains two of its inputs.
  \end{itemize}
  The first fact holds for $\Hset'$, and the second is~\eqref{eq:robust-count}.
  For the third, if a rounded point lies in $h^{+(L-i-1)\tau}$, then the exact Radon point lies in $h^{+(L-i)\tau}$, and \cref{lem:radon} applies.
  The output $z$ is not rounded.
  Hence $z\notin\bigcup\Hset'$ with probability at least $1-2\varphi/3$ given the outcome of \cref{lem:sample-ok}, and then $z$ is a $k_0'/\nu$-centerpoint of $S$ by \cref{lem:bad-family}.
  Finally, $k_0'/\nu\ge\beta-d/\nu\ge(1-\eps)\beta\ge(1+\eps)\beta/2$, since $\eps\le\frac13$, and \cref{lem:transfer} completes the proof.
\end{proof}

Scaling by $2^b$, each Radon computation of \cref{prop:bits} is a kernel computation for an integer matrix with entries of $O(b+B)$ bits.
Fraction-free Gaussian elimination~\cite{b-sime-68} performs it with $O(d^3)$ operations on integers of $O(d(b+B+\log d))=O(d^2(B+\log d)+d\log L)$ bits.
Hence the algorithm runs in polynomial time on a Turing machine, and its output is an exact centerpoint, not an approximation of one.

We expect the argument to extend to processes that create $N$ points one after the other, such as those of~\cite{cemst-acpir-96,hj-jcps-21,dp-ncptad-26}, with $N$ in place of $L$ and $O(\log N)$ additional bits.
There, the points alive after step $j$ would be measured against $h^{+(N-j)\tau}$.
A new point created at step $j$ lies in $h^{+(N-j)\tau}$ only if two (for Tverberg points, $s$) of its inputs lie in $h^{+(N-j+1)\tau}$, and a surviving point in $h^{+(N-j)\tau}$ already lay in the larger set $h^{+(N-j+1)\tau}$.
Making this precise requires checking the domination arguments of each of these analyses, and fitting the loss of $d$ points per halfspace into their quality, which we do not pursue.

The proof of \cref{prop:bits} applies without change to \textsc{LayeredTverberg}, under the same assumptions ($P\subset\mathbb{Z}^d$, $n\ge n_0$, and $\eps\le\frac13$): if the points of the layers are rounded, the output is a $(1-\eps)^2\beta$-centerpoint of $P$ with probability at least $1-\varphi$.
All coefficients of the linear programs then have $O(d(B+\log d)+\log L)$ bits.
Here every linear program must be solved exactly, that is, it must return a rational feasible point, for example by the ellipsoid method or by an interior-point method followed by rounding to an exact basic solution; an approximately feasible point need not have the halfspace property of \cref{lem:tverberg}.

\bibliography{refs}

\section*{AI Statement}
This paper was prepared with substantial help from the AI assistant Claude (Anthropic). It took part in developing the algorithms and proofs, checking the references, drawing the figures, and writing the text. The authors checked all results and are responsible for the content.




\end{document}